# A Survey Paper - Biometrics Employing Neural Network

Sajjad Bhuiyan
Department of Computer Science and Engineering, Syracuse University, Syracuse, NY, USA mhbhuiya@syr.edu

*Abstract-* Biometrics involves using unique human traits, both physical and behavioral, for the digital identification of individuals to provide access to systems, devices, or information. Within the field of computer science, it acts as a method for identifying and verifying individuals and controlling access. While the conventional method for personal authentication involves passwords, the vulnerability arises when passwords are compromised, allowing unauthorized access to sensitive actions. Biometric authentication presents a viable answer to this problem and is the most secure and user-friendly authentication method.

Today, fingerprints, iris and retina patterns, facial recognition, hand shapes, palm prints, and voice recognition are frequently used forms of biometrics. Despite the diverse nature of these biometric identifiers, the core objective remains consistent—ensuring security, recognizing authorized users, and rejecting impostors. Hence, it is crucial to determine accurately whether the characteristics belong to the rightful person. For systems to be effective and widely accepted, the error rate in recognition and verification must approach zero.

It is acknowledged that current biometric techniques, while advanced, are not infallible and require continuous improvement. A more refined classifier is deemed necessary to classify patterns accurately. Artificial Neural Networks, which simulate the human brain's operations, present themselves as a promising approach. The survey presented herein explores various biometric techniques based on neural networks, emphasizing the ongoing quest for enhanced accuracy and reliability. It concludes that The utilization of neural networks along with biometric features not only enhances accuracy but also contributes to overall better security.

## INTRODUCTION

Biometric recognition involves the utilization of distinctive physiological and behavioral characteristics, referred to as biometric identifiers, for the automatic recognition of individuals. Unlike traditional authentication methods that rely on possessions, biometrics verifies individuals based on inherent characteristics unique to them. These characteristics, or modalities, distinguish individuals within a population based on their innate physical or behavioral traits.

Despite the increasing deployment of biometric systems, these technologies encounter fundamental issues, including false matches and non-matches. It's important to note that a rejected user should not be automatically labeled as an imposter. The rejection might stem from the user experiencing difficulty with their biometric feature or the biometric system's inability to accurately process the individual's features.

With the increasing prevalence of technologies like e-banking, e-commerce, smart cards, home security, and smart cars, our world is becoming progressively digitized. Relying solely on passwords and keys is no longer sufficient for ensuring data security. Passwords can be a weak link in an organization's security system or an individual's privacy, as they are shareable, susceptible to theft, and can be cracked through various methods. This change has directed attention to biometric security as the sole effective method for confirming a person's identity.

Biometrics can achieve high precision using artificial neural networks (ANN). Biometric traits encompassing physical and behavioral characteristics such as fingerprints, face, iris, gait, and voice are inherent and unique to each individual. By employing different layers of mathematical processing, ANN enhances the accuracy and reliability of biometric security systems in verifying an individual's identity.

Since biometric traits are integral to an individual's being, applying ANN results in improvements, reducing susceptibility to errors and enhancing the system's intelligence, speed, and overall reliability.

## BIOMETRIC SYSTEM

Like human memory, a biometric system retains an object or individual's distinctive features or characteristics. Once information is stored in the system's memory, it can be compared to incoming details for identification and verification purposes. The primary applications of biometric systems include authentication, access control, and individual identification.

In contrast to traditional methods that rely on possession (e.g., keys or key cards) or knowledge (e.g., passwords or answers to security questions), biometrics employs physical features like facial structure, fingerprints, iris patterns, and vein maps, along with behavioral attributes such as voice, handwriting, or keystroke dynamics, for identification and access authorization. While Biometric Systems may seem complex, they operate through three straightforward steps: enrollment, storage, and comparison. Figure 1 below illustrates the sequence of these steps in the process [6].

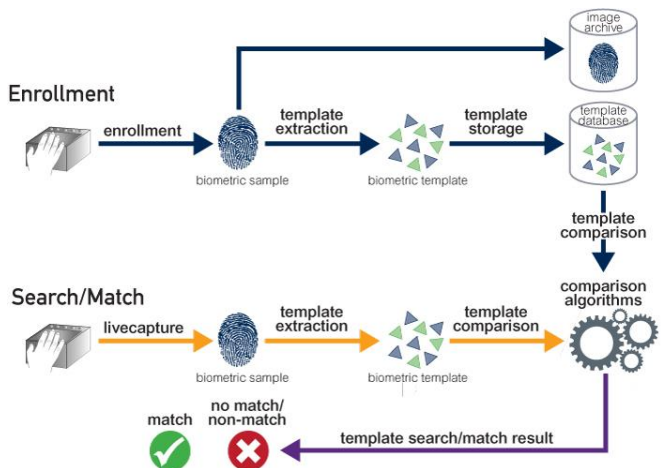

**Figure 1. 3 steps in Biometric Recognition**

The depicted image illustrates that the biometric system depends on various distinct stages: enrollment, live capture, template extraction, and template comparison. We can simplify the entire system into two main processes: enrollment and recognition. In enrollment, the system collects biometric samples, generates templates, and stores them for future comparisons. The system conducts a one-to-many comparison in the recognition process and identifies individuals by matching them with the stored templates.

During enrollment and live capture phases, template extraction transforms raw biometric data, like images or audio samples, into numerical digital templates. These templates are commonly created and saved at enrollment, facilitating quicker processing in subsequent comparisons. The comparison process involves algorithmic analysis to assess the similarity between two biometric templates. After this comparison, a match score is typically assigned. If this score exceeds a predefined threshold the user sets, the templates are deemed to match.

As biometric authentication relies on physiological or psychological characteristics, there is no need to carry keys or documents for identification. It is a fast, hassle-free, reliable, and cost-effective method.

**Some Biometrics techniques and their pros and cons:**

Various biometric techniques are utilized across applications, from smartphones and banking systems to home security. Much research has been conducted on biometrics technologies, and several specific technologies are already in use in real-time security systems or applications. Some of the popular and deployable biometrics available today include Face recognition, Fingerprint scanning, Iris recognition, Voice recognition, Hand geometry, Palm print recognition, Palm vein recognition, Ear recognition, Retina scanning, Gait analysis, Signature recognition, and Handwriting analysis [7].

| Biometric trait | Main advantage | Defect | Security level | Sensor |
|---|---|---|---|---|
| Voice | Natural and convenient | Noise | Normal | Noncontact |
| Face | Remote capture | Lighting conditions | Normal | Noncontact |
| Fingerprint | Widely applied | Skin | Good | Contact |
| Iris | High precision | Glasses | Excellent | Noncontact |
| Finger vein | High security level | Few | Excellent | Noncontact |

**Table 1: Overview of features in current biometric systems [2].**

Table 1 above lists popular biometric systems along with their advantages, defects, level of security, and cost. While biometrics appears promising, it is essential to comprehend both its advantages and disadvantages.

A biometric match is typically not 100% accurate, and a typical biometric system has two types of errors. A false reject (FR) error occurs when an authorized person is incorrectly rejected during the authentication process. Conversely, a false accept (FA) error happens when an unauthorized person is wrongly accepted, gaining access under false pretenses. These two kinds of errors have an inverse relationship and are typically managed through a concept called a threshold.

The threshold in biometric matching is the level where it's reliably determined that a biometric sample corresponds to a specific reference template. To enhance the system's security, the threshold can be adjusted, either increased or decreased.

The administrator will select a measure of similarity (threshold) upon which the system will determine whether there is a match or not. Choosing small threshold results in high security, but a lower threshold range reduces security. Higher security increases the likelihood of a false rejection rate, while lower security increases the risk of a false acceptance rate. Computer scientists are diligently working to minimize both rates. Now, let's examine some of the advantages and drawbacks of biometric systems.

**Some of the Advantages of Biometric System**
- Improved security
- Improved customer experience
- Elimination of forgotten or stolen passwords.
- Accurate and positive
- Reduced operation costs
- Offers modality
- Very difficult to forge

Acts as a non-transferable key.

**Some of the drawbacks of the Biometric System [16]**
- Physical Traits Are Not Changeable
- Due to the Error Rate, 100% accuracy is not possible
- Delay
- Complexity
- Unhygienic
- Scanning Difficulty
- Physical Disability
- Environment and Usage can affect measurements
- Additional Hardware Integration

## BIOMETRIC TECHNIQUES

In the contemporary world, insecurity is a significant issue in various sectors, and we are employing multiple computer techniques to address this concern. Traditional verification methods, such as PINs, passwords, and critical badges, are well-known. The advantage of biometric features lies in their universality, measurability, uniqueness, security, and permanence.

As observed earlier, various biometric techniques are accessible. While some are still in the research stages, many technologies are available for real-time applications. Face recognition biometrics, for instance, is widely utilized in systems like Uber, mobile phones, and more. It stands out as one of the mainstream biometric authentication methods [4]. Face biometrics is particularly convenient due to its non-intrusive nature. Furthermore, commercially accessible types of biometric authentication encompass fingerprint, finger geometry, hand geometry, palm print, iris pattern, retina pattern, facial recognition, voice comparison, and signature dynamics.

We are aware that biometrics is highly reliable, convenient, and secure. While achieving a 100% accuracy match in biometrics is not currently feasible, it enables the identification and authentication of individuals based on distinctive and verifiable data. In the authentication process, the person's data is compared with a stored template to ascertain resemblance. As depicted in Figure 2, a conventional flow graph can briefly summarize the biometric authentication process for a specific pattern.

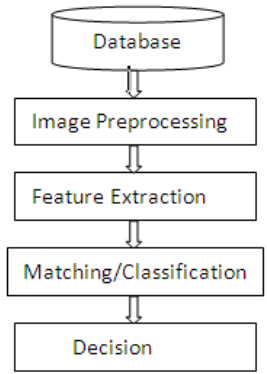

**Figure 2. Conceptual Overview Flow Graph.**

The conceptual overview involves the following steps:
1) Image Acquisition (from database): In this stage, the image of a person is obtained and processed. This serves as the fundamental first step, and once the image is acquired, several processing steps are applied.
2) Image Pre-processing: In this stage, the image undergoes further processing, including conversion to grayscale, application of thresholds, quality assessment, noise removal, and edge detection.
3) Feature Extraction: Feature extraction involves obtaining more relevant data from the original image. The primary purpose of this step is to reduce the data while preserving the excellent quality of the original. The result is a more compact and relevant product.
- Matching/Classification: The final stage involves matching and classification to grant appropriate access. Classification entails grouping things based on their characteristics. Once classified, the authorization process becomes faster and easier.
- In summarizing biometric approaches, it's important to note that a biometric system operates in two fundamental modes: identification and verification. In identification mode, the system aims to identify a person by analyzing a biometric pattern derived from their unique biometric traits.
- In the verification phase, the system compares the pattern being verified exclusively with the individual's distinct template. It assesses whether the degree of resemblance between the pattern and the template is sufficient to authorize entry into the protected system or area.

A threshold is utilized to reduce instances of incorrect acceptances or denials. The degree of similarity between a biometric pattern and a template is measured by scores, also known as weights. A higher score signifies a closer match. As previously described, access to the system is only permitted if the score of an identified individual (in identification mode) or the score in the comparison against the pattern (in verification mode) exceeds a predetermined threshold level.

## TYPES OF BIOMETRICS

Biometrics stands out as the foremost means of identifying and authenticating individuals, offering reliability and speed by leveraging unique biological characteristics. Behavioral biometrics involves measuring and identifying distinctive patterns related to human activities, such as keystroke dynamics, gait analysis, and voice ID. On the other hand, physiological biometrics pertains to specific body shapes, measurements, dimensions, and characteristics. This category encompasses fingerprints, hand shape, vein patterns, the eyes (iris and retina), and facial features. The realm of biometric methods extends beyond common knowledge, and we will categorize them into different groups as follows:

**Visual:**
EAR: In biometrics, an individual can be identified using the shape of the ear.
Eyes - Iris Recognition: The iris in each eye is distinct and can serve as a biometric identifier for a person.
Eyes - Retina Recognition: The unique vein patterns located at the back of the eye can be used for personal identification.
Face Recognition: Facial analysis involves examining features or patterns to authenticate or identify a person.
Fingerprint Recognition: Fingerprint recognition technology operates by identifying the ridges and valleys present on human fingertip surfaces to verify a person's identity.
- Finger Geometry Recognition: Finger geometry recognition involves using the three-dimensional shape of the finger for identity verification.
- Hand Geometry Recognition: The geometric features of the hand are used to determine identity.
• Signature Recognition: Signature recognition authenticates an individual through analyzing handwriting style.

**Behavioral:**
- Gait: Gait is considered a component of behavioral biometrics as one's walking style or gait determines identity.
- Typing Recognition: Recognition based on the behavior in one's typing style can identify that person.
- **Chemical:**
  - DNA: Involves the identification of an individual through the analysis of DNA segments.

  **Vein:**
  - Vein Recognition: The patterns of veins in the human finger or palm can serve as biometrics for identifying an individual.

  **Olfactory:**
  - Odor: human odor is also unique to a person and can be used for biometric identification

  **Auditory:**
  - Voice - Speaker identification: The speaker's voice can be compared against the known template to verify identity.

Today's widely used deployable biometrics include Face, Fingerprints, Iris, Voice, Hand geometry, Palm print, Palm veins, Ear, Retina, Gait, Signature, and Handwriting [7]. The further investigation in this paper will focus predominantly on neural networks. Our review is based on the broad concept of applying neural networks across different biometric methodologies.

# NEURAL NETWORKS

An Artificial Neural Network (ANN) is a model for processing information that draws inspiration from the workings of the biological nervous system, especially the brain. It is a computational system that emulates the human brain's processing methods, structure, and learning capabilities. An artificial neural network (ANN) is a human brain simulation. The natural neural network in the brain can learn new information and adapt to a dynamic environment. Figure 3 illustrates the basic structure of a simple Artificial Neural Network.

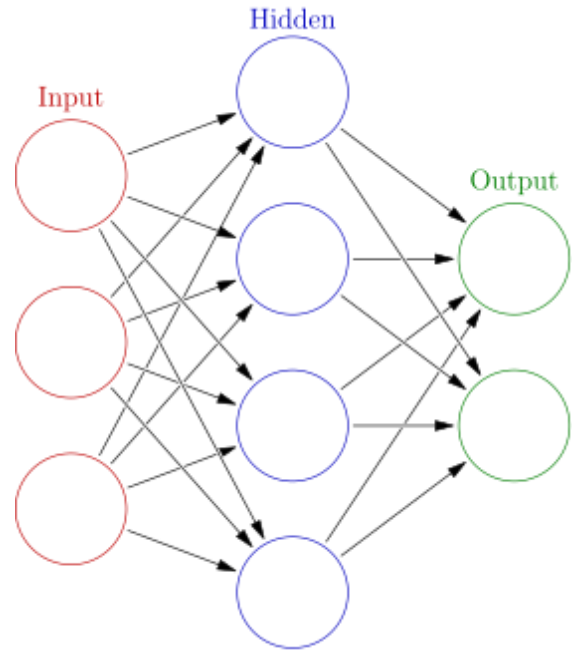

**Figure 3. Artificial Neural Network**

An artificial neural network is a computational model consisting of interconnected nodes. The inspiration behind the artificial neural network (ANN) lies in simplifying the neurons found in the human brain. Figure 3 depicts each circular node as an artificial neuron, with arrows showing connections from one artificial neuron's output to another's input.

The Artificial Neural Network (ANN) differs from a typical computer program in several ways. Its distinctive features include adaptive learning, self-organization, parallel operation, and fault tolerance, among others [13]. In comparison to the human brain, the ANN demonstrates notable speed.

The Artificial Neural Network (ANN) tackles extensive parallel processing, distributed data representation, learning capacity, generalization capacity, and fault resilience. It proves highly effective in tasks such as pattern classification, clustering, categorization, function approximation, prediction and forecasting, optimization, and control. The ANN is gaining prominence across various applications, including pattern recognition, weather prediction, handwriting recognition, face recognition, autopilot, robotics, and more [1].

The Artificial Neural Network (ANN), a connectionist system, learns to execute tasks by analyzing examples. It can perform tasks based on examples without the need for explicit programming of task-specific rules. For example, in the case of image recognition, an ANN can learn to recognize a dog by examining images that are manually categorized as either "dog" or "not dog." The acquired knowledge can then be applied to identify dogs in other images. The ANN system can make predictions and identifications based on the examples it has been exposed to.

# WHY USE NEURAL NETWORKS?

An artificial neural network (ANN) is a computational framework modeled after the architecture and functionalities of biological neural networks. Below are several reasons for employing neural networks:
- Neural networks can derive meaning from complicated data.
- A neural network is characterized by adaptive learning capabilities, allowing it to learn how to perform tasks based on the data provided during training.
- An ANN can develop its own structure or representation of the information it acquires during the learning phase.
- A considerable number technologies have already been developed and tested and are available in the commercial market.
- ANNs are widely used in pattern recognition because they generalize and respond to unexpected inputs.
- Neural Networks prediction accuracy is generally high
- Neural Networks can work with a large number of variables or parameters

In practical scenarios, many relationships between inputs and outputs are nonlinear and intricate. Neural Networks are capable of learning and modeling these complex, nonlinear relationships.

In summary, Neural Networks are straightforward mathematical models that take a set of vectors as input and generate a set of vectors as output after performing a series of operations.

We currently employ them because they can be easily combined and stacked to yield improved results in a concept known as Deep Learning. This is feasible due to the computational power available to us. Neural Networks are robust and versatile systems, embodying a hybrid approach.

# RELATED WORK

Now that we understand how biometric systems function and what Neural Networks are, we will explore some biometric-related work that employs Neural Networks. While there are numerous research papers available, we will focus our discussion on the following commonly used biometric systems that utilize Neural Networks as a classifier:
- Face Recognition/Detection
- Fingerprint Recognition
- Finger-vein biometric identification
- Irish Recognition
- Gait Recognition

These research papers aim to investigate how Neural Networks can enhance the reliability and effectiveness of the biometric process.

### Face Recognition/Detection:

Face recognition is a widely used and significant method of biometric authentication. In this process, an image of a face is compared with those stored in a database. Access is granted once a match is found. Although face recognition systems are nearing human-level accuracy in identifying individuals across diverse and challenging datasets, they are still vulnerable to presentation attacks (PA), commonly known as spoofing attacks. An uncomplicated action, like showing a printed photograph, can trick a face recognition system that lacks adequate protection. Many strategies have been devised to counter presentation attacks, yet most face difficulties in identifying advanced attacks, like those using silicone masks. With the continuous enhancement of presentation attack instruments, ensuring dependable detection of such attacks through visual spectra alone remains a significant challenge.

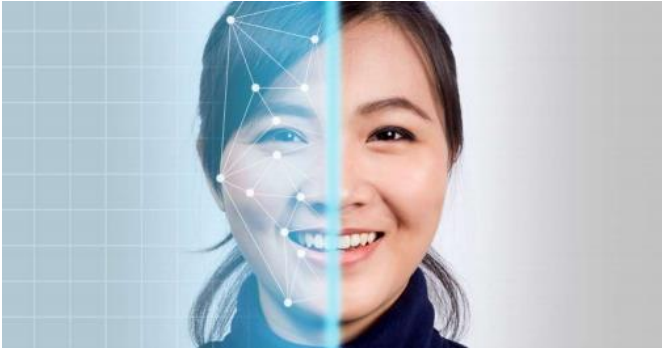

Figure 4, Abstract human face into features

Conventional face recognition algorithms detect facial characteristics, extracting landmarks or features from a subject's face image. For example, as shown in Figure 4, an algorithm may assess the relative position, size, and contours of the eyes, nose, cheekbones, and jaw. These identified features are subsequently used to find other images with similar attributes. Such algorithms can be complex, requiring significant computational power, and, as a result, may exhibit slow performance. Additionally, they can be inaccurate, particularly when faces display pronounced emotional expressions, as the landmarks' size and position can substantially alter in these scenarios. As previously mentioned, these biometric systems are susceptible to presentation attacks. Detecting such attacks becomes increasingly complex, especially in realistic 3D and partial attack cases.

In their paper proposing a face recognition system using an artificial neural network, Shahrin, Nazaruddin, and Marzuki describe two distinct phases: enrollment and recognition/verification, as illustrated in Figure 5. Initially, the paper discusses face recognition methods as a typical pattern recognition problem. The text then contrasts newer methods centered on representation and recognition with their suggested system, which similarly prioritizes representation and recognition through artificial neural networks. This system consists of multiple components: Image Acquisition, Face Detection, Training, Recognition, and Verification.

In their paper, the authors initially detail the proposed face recognition system, discussing the methodology employed. Subsequently, the paper evaluates the system's performance. This method employs histogram equalization and homomorphic filtering techniques and contrasts their outcomes with those obtained using Euclidean Distance and Normalized Correlation classifiers. The system demonstrates promising results in both face verification and recognition tasks. A visual overview of the face recognition system is provided in the following diagram.

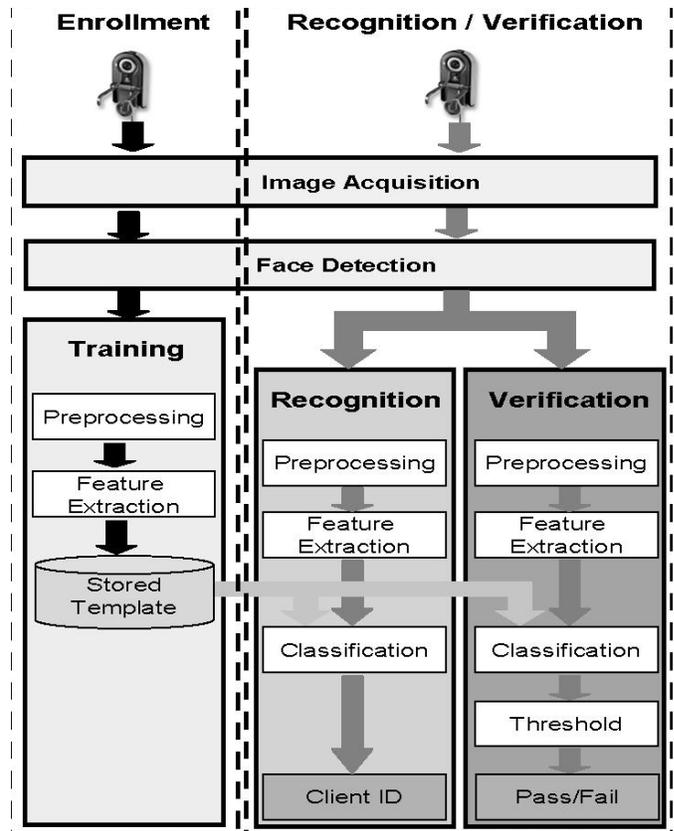

**Figure 5. Block Diagram for the Face Recognition System [5]**

The block diagram illustrates that the model comprises multiple modules: The system encompasses Image Acquisition, Face Detection, Training, Recognition, and Verification. In this structure, the goal of the classification sub-module is to map the feature space of the test data to a discrete set of labeled data, serving as a template.

The classification methods used are Artificial Neural Networks, Euclidean Distance, and Normalized Correlation [5]. The feature extraction process employs algorithms such as Principal Component Analysis (PCA) and Linear Discriminant Analysis (LDA).

The ANN classifier is advantageous for classification due to its remarkable generalization capabilities and superior learning capacity. In this context, the ANN receives a feature vector as input and trains the network to master a complex mapping for classification, thereby eliminating the necessity to simplify the classifier. ANN has also been applied in the context of face recognition.

The experimental findings derive from ten facial images for each of the 20 subjects. The papers deduce that the face verification neural network classifier can be regarded as the most effective among the three classifiers, as it consistently delivers strong performance across all experiments utilizing both PCA and LDA for feature extraction.

| Feature Extractor | Classifier | FAR = FRR (%) FAR | FAR = FRR (%) FRR | HTER (%) |
|---|---|---|---|---|
| PCA | E.D. | 6.540 | 12.590 | 9.565 |
|  | N.C. | 5.690 | 5.560 | 5.625 |
|  | N.N | 4.140 | 3.700 | 3.920 |
| LDA | E.D. | 6.030 | 6.300 | 6.165 |
|  | N.C. | 3.660 | 4.810 | 4.235 |
|  | N.N | 4.660 | 5.190 | 4.925 |

Table 2. Results of Face Verification Utilizing Homomorphic Filtering and Histogram Equalization

Regarding face recognition, the paper determines that the neural network (N.N) classifier yields the highest accuracy when combined with homomorphic filtering and histogram equalization, mainly when using the PCA feature extractor. Conversely, when paired with the LDA feature extractor, the normalized correlation (N.C.) classifier achieves the highest accuracy.

| Feature Extractor | Classifier | Recognition (%) |
|---|---|---|
| PCA | E.D. | 91.85 |
|  | N.C. | 91.85 |
|  | N.N | 92.59 |
| LDA | E.D. | 90.00 |
|  | N.C. | 92.22 |
|  | N.N | 85.56 |

Table 3. Recognition Results using Homomorphic Filtering and Histogram Equalization

In summary, the paper introduces a face recognition system that utilizes artificial neural networks for face verification and face recognition tasks, employing photometric normalization for comparative analysis. The experimental findings indicate that the neural network (N.N.) outperforms both Euclidean distance and normalized correlation decision rules regarding overall verification performance when using both PCA and LDA feature extractors.

**Fingerprint Recognition**
Fingerprints have been used for identification for a very long time. It is also one of the oldest of all the biometrics. In fingerprint identification, the system will recognize three basic patterns that we all humans have:
- Arch – ridges and valleys form an arch-like pattern in this pattern. This is the rarest type of fingerprint. This pattern lacks cores, lines, or deltas, which makes it unique. About 5 percent of the population has this type of pattern.
- Loop – The loop pattern is the most prevalent type of fingerprint, observed in about 60-70 percent of the population. In this pattern, the ridges enter from one side of a finger, form a loop, and then exit from the same side.
- Whorl – Approximately 25 percent of the population exhibits this fingerprint pattern, where ridges and valleys create a circular formation around a central point on the finger, resembling a whorl.

Fingerprint recognition has become one of the most prevalent authentication methods in security applications today. This technology is extensively implemented across diverse fields, including government, corporate organizations, libraries, universities, and banks. Its popularity in biometric systems, over alternatives like iris, face, hand, voice, and signature recognition, stems from its unique and distinctive nature. Traditional techniques are less effective in analyzing fingerprint texture features than neural network classifiers [3].

In their research, K. Martin, J. Winston, N. Ponraj, Y. JC, E Jeba, and A. Clara [3] explore and categorize fingerprint recognition using Euclidean distance and neural network (NN) classifiers to achieve enhanced accuracy. Minutiae of the fingerprint provides the most accurate results compared to the ridge shape features since the ridge shape features contain only the loop (∩), arch (Δ), and whorl(o) information, which may lead to false recognition. Minutiae details are usually precise. Minutiae details consist of two minutiae features: Ridge ending and Bifurcation.

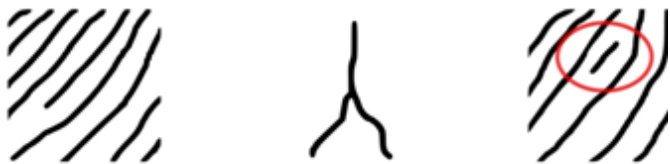

Figure 6 illustrates minutiae features, including a) Ridge ending, b) Bifurcation, and c) Short ridge.

The paper describes how pre-processing primarily focuses on the manipulation of low-intensity images. Pre-processing methods address redundancies like scars, extreme dryness, moisture, dust, varying pressure, and more. The primary goal of these pre-processing techniques is to enhance the informative content of images by reducing unwanted distortions and amplifying essential image features.

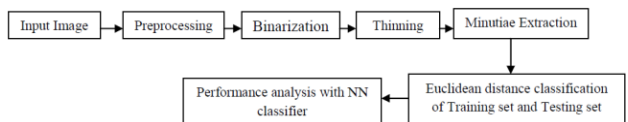

Figure 7. Overview of Fingerprint recognition system

The preprocessing technique's primary aim is to enrich images' informational quality by minimizing unwanted distortions and improving image characteristics. The image enhancement approach encompasses two key processes aimed at noise reduction and improved image capture: Histogram Equalization and Fast Fourier Transform.

The paper describes how the Artificial Neural Network (ANN) takes inspiration from brain processes to create algorithms that can adeptly model complex patterns and prediction problems. Within this framework, a neural network is one of the various machine learning algorithms capable of addressing classification challenges. Images are both trained and tested in this classification process.

The experiment was conducted using 80 images, and the results were analyzed. The Euclidean distance classifier performed effectively across all 80 images. The resulting values were then processed through a neural network tool for additional classification. This method demonstrated improved computational time, reduced error rates, and enhanced accuracy. The outcomes were compared with other methods, as shown in the table.

| Methods | Accuracy | MSE(Mean square error) % |
|---|---|---|
| Proposed method | 99% | 0.01 |
| P.Gnanasivam | 94% | - |
| Feng Zhao | 78% | 0.02 |
| Mouad.M.H.Ali | 80.3% | 0.2 |
| Lu Jiang | 94% | 4.97 |

Table 4. Comparison with the other methods

The table shows the experiment's outcome and compares accuracy with another method. Their proposed method shows 99% accuracy with a mean square error of 0.01%.

The paper concludes that matching the input and test images is challenging, although the query image is enrolled at a different angle. The Euclidean distance and neural network classifier analyzes the test and input images most accurately.

**Finger-vein biometric identification:**
Fingerprint authentication and face recognition are commonly used biometric authentications. The finger vein recognition recently received much attention. The finger vein authentication accuracy is high compared to other biometric technologies. Finger vein recognition employs pattern recognition methods, utilizing images of vein patterns found beneath the surface of human fingers. In contrast to biometrics like fingerprints and facial recognition, vein patterns are internal, rendering them nearly impossible to duplicate. This attribute makes finger vein biometrics a more secure alternative, less vulnerable to forgery, damage, or alterations over time.

Ensuring the security of personal belongings, privacy, and information is increasingly crucial. Many traditional biometric identification systems, relying on physiological traits and behavioral patterns like faces, irises, or fingerprints, come with various limitations and susceptibility to spoofing.

Traditional finger-vein recognition approaches require significant image processing to remove noise and extract and amplify features before image classification to achieve high accuracy. A distinct advantage of convolutional neural networks (CNNs) compared to these conventional methods is their ability to perform feature extraction, data dimensionality reduction, and classification concurrently within one network framework [3].

In their study, Sayafeeza R., Mohamed K, and Rabia [2] focused on utilizing a convolutional neural network (CNN) for finger-vein biometric identification. A key benefit of using CNNs compared to traditional methods is their capacity to concurrently perform feature extraction, data dimensionality reduction, and classification within a single network architecture. Moreover, this approach demands minimal image preprocessing because the CNN is resilient to noise and minor misalignments in the captured images.

The paper by Sayafeeza R., Mohamed K, and Rabia [2] introduces a simplified four-layer CNN featuring a combined convolutional-subsampling structure specifically for finger-vein recognition. They adapted and implemented the stochastic diagonal Levenberg-Marquardt algorithm to train the network, leading to a quicker convergence time.

While finger vein biometrics offer numerous benefits, they also face practical challenges, such as dependency on the quality of the image acquisition device and the occurrence of optical blurring in captured images. Factors like inadequate lighting and the quality of the capturing device can lead to images being excessively dark or bright, misaligned, and other issues.

Traditional finger-vein recognition techniques utilize intricate image-processing algorithms to address these challenges. However, recent studies indicate that computational intelligence (CI) methods can effectively handle these issues, encompassing neural networks, fuzzy logic, and evolutionary computing.

Figure 8 illustrates the suggested CNN structure for finger-vein biometric recognition. This CNN comprises four layers, labeled as C1, C2, C3, and C4 convolutional layers, excluding the input layer. The architecture is referred to in the paper as the 5-13-50 model, indicating the presence of 5, 13, and 50 feature maps in the C1, C2, and C3 layers, respectively.

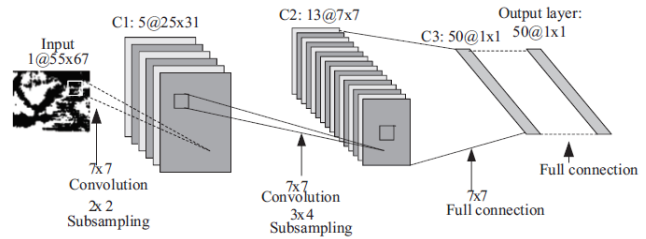

Figure 8. Proposed CNN architecture

The proposed CNN was evaluated using an internally developed finger-vein database comprising 50 subjects, each contributing ten samples from their fingers. According to the paper, the system attained a 100.00% identification rate when dividing the samples into 80% for training and 20% for testing. The system was further tested with more subjects; for 81 subjects, it achieved an accuracy of 99.38%.

| Number of subjects | Number of training samples | Number of test samples | Accuracy (%) |
|---|---|---|---|
| 50 | 400 | 100 | 100.00 |
| 81 | 648 | 162 | 99.38 |

Table 5. The attained accuracy levels correspond to datasets involving 50 and 81 subjects, respectively.

The paper introduces a novel method for finger-vein biometric identification using a CNN. The experimental results indicate that the proposed CNN-based approach, which includes preprocessing but omits the resource-intensive segmentation (local dynamic thresholding) process, achieves optimal accuracy. Furthermore, the paper concludes that this method does not necessitate segmentation processes or noise filtering. For future work, the proposed CNN model can be evaluated with a more complex database encompassing more subjects.

### Irish Recognition:

Irish recognition is exciting, unique, and secure. Irish recognition has various advantages, such as speed, simplicity, and accuracy. Irish biometrics are unique; left and right Irish patterns are different. It is also challenging and risky to alter or tamper with the Irish patterns.

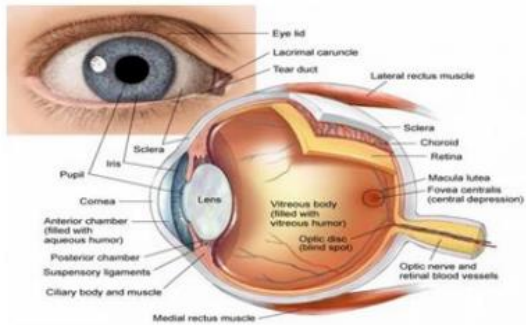

Figure 8b: Iris Structure

The iris possesses numerous unique characteristics, including arching ligaments, furrows, ridges, crypts, corona, and freckles. Additionally, it is shielded from external elements by the cornea and eyelids. The delicate radial patterns of the iris remain constant and unchanging from around one year of age throughout a person's life. These factors make iris recognition a promising area in biometrics [12]. The iris can function as a "living passport" or "living password." The basic model of an iris recognition system is depicted in the subsequent figure.

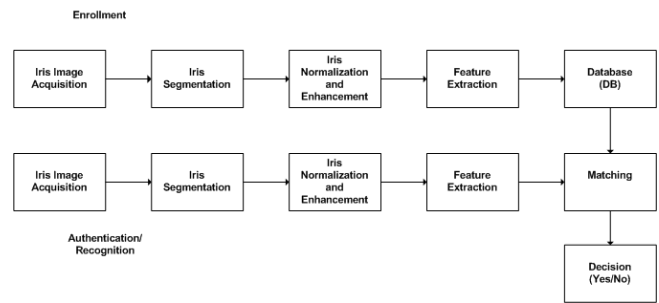

Figure 9. Generalized Block Diagram of Iris Recognition System.

The iris recognition procedure involves capturing an image, pre-processing, and identifying the iris region in the eye image. The pre-processing of the iris image encompasses localization. Typically, the identification process begins with capturing an eye image from a person using a digital camera. During the Enrollment stage, this image is stored in the database following segmentation, enhancement, and feature extraction. In the Authentication and Recognition stage, the newly captured image undergoes feature extraction and is then compared with the one stored in the database to make a decision.

In their research paper, V. Saishanumuga Raja and S.P. Rajagopalan [12] propose a method for personal identification based on iris recognition utilizing genetic algorithms and neural networks. The iris recognition process involves the localization of the iris region and the generation of a dataset of iris images, followed by iris pattern recognition. A neural network is employed to address issues such as a low recognition rate, low accuracy, and increased recovery time. The paper indicates that the genetic algorithm is utilized to optimize the parameters of the neural network. The simulation results demonstrate a commendable identification rate and reduced training time.

The paper highlights the importance of neural networks' capacity to learn and identify relationships between objects and patterns relevant to the real world, thus exhibiting helpful behavior. In this regard, neural networks are valuable tools for tackling complex problems. The Artificial Neural Network (ANN) is a versatile computational algorithm that can develop complex hypotheses to explain significant correlations between features without needing prior knowledge from the dataset.

Their research developed a neural network that uses iris images as input. The neural network's hidden layer nodes are optimized using a genetic algorithm, enabling them to learn from the inputs provided. During this operation, the Artificial Neural Network matches the scanned iris image against a stored image for personal identification. For successful identification, the network's nodes must learn and remember the iris features each time an image is fed into the system. The study's objective is to increase the learning rate of the Neural Network and enhance its accuracy in individual identification. The results of their experiment are displayed in Table 6.

| Methodology | Accuracy Rate | Average time |
|---|---|---|
| Daugman [5] | 100% | 90 s |
| Boles [12] | 92.64% | 110 s |
| Neural network without GA | 93.3 % | 20 s |
| Neural network with GA | 98.48% | 10.8 s |

Table 6. Recognition performance comparison with the existing methods

The paper determines that the suggested approach, which employs a genetic algorithm to optimize the current neural network methodology, achieves higher accuracy and shorter learning times. The findings indicate that this method significantly improves accuracy in iris recognition and reduces the error rate. According to the paper, when optimized by the genetic algorithm, the proposed method attains superior accuracy compared to the conventional neural network approach.

### Artificial Neural Network Based Gait Recognition

The precision of gait recognition holds considerable importance, particularly in industrial and consumer applications. It has significant applications in video surveillance, virtual reality, online gaming, medical rehabilitation, collaborative space exploration, and more. A.S.M. Hossain Bari and Marina L. Gavrilova [18] have suggested a novel architecture that utilizes a deep learning neural network for highly accurate, rapid, and cost-efficient gait recognition based on Kinect technology. They have introduced two innovative geometric features: joint relative cosine dissimilarity and joint relative triangle area, which are view-independent and invariant to changes.

The incorporation of these features notably enhances the performance of recognition. The proposed neural network model undergoes training with a feature vector that includes dynamic joint relative cosine dissimilarity and joint relative triangle area. The use of the Adam optimization method aids in progressively reducing the loss of the objective function. In their research, the efficiency of the proposed deep learning neural network architecture is evaluated using two publicly available 3D skeleton-based gait datasets recorded with the Microsoft Kinect sensor.

The accuracy, precision, recall, and F-score of the proposed neural network architecture have been confirmed through experimental validation. When trained with the newly introduced dynamic geometric features, this architecture surpasses other prominent Kinect skeleton-based gait recognition techniques in performance.

The paper underscores that gait recognition is widely used in various industrial and consumer settings, such as video surveillance, virtual reality, online gaming, medical rehabilitation, and collaborative space exploration. As a behavioral biometric, Gait is a favored method for discreet, remote authentication. Since gait biometrics are captured from a distance, it allows for remote identification of individuals. Consequently, gait recognition systems find applications in multiple domains such as personal authentication, human action recognition, gender determination, surveillance, analysis of atypical gaits for medical purposes, and assessing psychological states. Gait analysis fundamentally relies on the motion of different body joints. Although factors like physical injuries and fatigue can affect a person's gait, it remains challenging to replicate another individual's walking pattern, as noted in the study [18]. This unique aspect renders biometric gait particularly effective for robust and unobtrusive authentication systems, enabling the authentication of subjects without the need for direct interaction.

Additionally, the paper emphasizes that deep learning, a prominent machine learning (ML) approach, presents new opportunities for sophisticated human motion analysis. This contrasts with traditional machine learning methods, which have previously been used for gait recognition. While conventional techniques have yielded satisfactory outcomes with Kinect-based gait sequences, they are hindered by notable limitations.

Their paper introduces a novel deep-learning neural network designed explicitly for Kinect-based gait recognition. The architecture of this proposed deep-learning neural network is depicted in Figure 10.

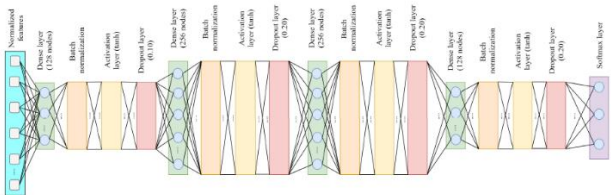

Figure 10. Architecture of the Suggested Deep Learning Neural Network.

The proposed neural network architecture inputs standardized features, specifically the joint relative triangle area and joint relative cosine variation. During a pre-processing phase, the feature vectors undergo standard normalization to mitigate the impact of outliers on the training of the neural network model. Identification labels for each individual are converted into a one-hot encoded format. Subsequently, these normalized feature vectors and one-hot encoded identification labels are inputted into the hidden layers to train the deep learning neural network.

Table 7 compares the recognition performance of the suggested gait recognition method with current leading techniques to demonstrate the efficacy of this approach.

| Gait recognition methods | Accuracy | Precision | Recall | F-Score |
|---|---|---|---|---|
| Ball et al. [10] | 37.55 | 37.84 | 38.11 | 34.25 |
| Preis et al. [11] | 75.46 | 77.70 | 75.34 | 73.71 |
| Sun et al. [9] | 79.76 | 80.12 | 79.32 | 75.74 |
| JRA + JRD [12] + Proposed DLNN + tanh + SGD | 91.28 | 91.07 | 92.15 | 90.19 |
| JRA + JRD [12] + Proposed DLNN + tanh + RMSProp | 91.93 | 91.09 | 91.71 | 90.06 |
| Yang et al. [13] | 94.88 | 94.67 | 95.02 | 93.92 |
| JRA + JRD [12] + Proposed DLNN + tanh + Adam | 95.62 | 95.92 | 95.94 | 95.14 |
| **Proposed features + Proposed DLNN + tanh + Adam** | **98.08** | **98.0** | **98.26** | **97.81** |

Table 7: Comparative Analysis of Gait Recognition Techniques Using the Gait Biometry Dataset.

The paper concludes by presenting table results, showing that the accuracy of the method by Ball et al. surpasses that of Preis et al. by 17.30% and Sun et al. by 12.63%. Moreover, the process they propose attains a recognition accuracy of 8.63%, higher than the one achieved by Yang et al.

## CONCLUSION

Various types of biometrics each have their advantages. After reviewing multiple research papers, it can be concluded that integrating neural networks with biometric features offers enhanced security compared to other techniques. Further studies in this work will focus more specifically on a biometric technique that has garnered attention, demonstrating improved accuracy and performance when combined with neural networks. The utilization of neural networks along with biometric features not only enhances accuracy but also contributes to overall better security.